\documentclass[a4paper,10pt,twocolumn]{article} 
\usepackage{graphicx}
\usepackage{subfigure}
\usepackage{amsmath}
\usepackage{amssymb}
\usepackage{wrapfig}
\usepackage{color}
\usepackage[english]{babel}
\usepackage[warning,math]{easyeqn}
\usepackage[thinlines]{easybmat}
\usepackage{enumerate} 
\usepackage{multirow} 
\usepackage[pdftex,
            pdfauthor={Yves-Henri Sanejouand},
            pdftitle={On the unknown proteins of eukaryotic proteomes},
            pdfsubject={Singletons},
            pdfkeywords={Ubiquitous, unknown, singleton proteins},
            pdfproducer={Latex with hyperref},
            pdfcreator={Texmaker}]{hyperref}
\begin{document}

\title{\textbf{On the unknown proteins of eukaryotic proteomes}}

\author{Yves-Henri Sanejouand\footnote{yves-henri.sanejouand@univ-nantes.fr}\\ \\
		US2B, UMR 6286 of CNRS,
        Nantes University, France.} 
\date{September 22$^{th}$, 2022}
\maketitle

\section*{Abstract}

In order to study unknown proteins on a large scale,
a reference system has been set up for the three major eukaryotic lineages,
built with 36 proteomes as taxonomically diverse as possible.
Proteins from 362 eukaryotic proteomes with no known homologue in this set were then analyzed, focusing noteworthy on singletons, that is, on unknown proteins with no known homologue in their own proteome.

Consistently, 
according to Uniprot, for a given species,
no more than 12\% of the singletons thus found are known at the protein level.
Also,
since they rely on the information
found in the alignment of homologous sequences,
predictions of AlphaFold2 for their tridimensional structure are usually poor.

In the case of metazoan species, the
number of singletons seems to increase as a function of the evolutionary distance from the reference system.
Interestingly, no such trend is found in the cases of
viridiplantae and fungi, as if
the timescale on which singletons are added to proteomes were different in metazoa and in other eukaryotic kingdoms.
In order to confirm this phenomenon, further studies of proteomes
closer to those of the reference system are however needed.

\vskip 1cm
\textbf{Keywords}: Ubiquitous proteins, Singletons, Metazoa, Viridiplantae, Fungi, Evolutionary distance, Uniprot, AlphaFold2.

\section*{Introduction}

Since the earliest genome sequencing projects proteins with no known homologue have been found in significant amounts \cite{Fischer:03,Tautz:11}. Various origins have been proposed for such unknown proteins \cite{Wang:03,Mclysaght:20}: gene duplication \cite{Ohno:70,Ohta:89}, 
followed by a neutral drift of their sequence \cite{Jukes:69,Kimura:83,Trinquier:99,Tawfik:08},
incorporation of transposable elements \cite{Maralba:09,Kaessmann:16},
\textit{de novo} genesis, that is, 
evolution from random amino acid sequences \cite{Jacobs:93,Mclysaght:09,Tautz:09,Vidal:12,Bornberg:17}, \textit{etc}.

With the advent of massive genome sequencing projects \cite{Detter:09,Chinamap:20,Guo:21}, the total number of unknown proteins is expected to increase dramatically. On the other hand, the definition of what is an unknown protein relies on the information already available
which, in the case of eukaryotic species,
is biased towards what is known for a small set of model species, as well as for species close to, or the most useful for the human one.
As a consequence, a robust definition 
is needed for unknown proteins, especially in order to perform quantitative comparisons between species. 

On the other hand, in a context of an ongoing massive extinction of species \cite{Ferrer:11,Todd:15},
it could prove worthwhile focusing the efforts 
on preserving those (their genomes, at least) that are
the more likely to prove useful for humans in the future
\cite{Crozier:97,Donoghue:10,Small:11}. 
In particular, 
molecules of particular importance for our health
have been found in various species \cite{Rates:01,Shen:16,Gwozdzinski:18}.
Since such molecules are synthesized by enzymes
with original specificities (or functions),
the hypothesis that
species hosting a lot of unknown proteins
may prove more likely to yield
enzymes with promizing characteristics needs to be considered. 

Herein, as a first step towards this end,
an instrumental definition is proposed for unknown proteins, based on the setup of a reference system for the proteomes of the three major eukaryotic kingdoms, namely, metazoa, viridiplantae (land plants) and fungi. While with such a definition the status of a given
protein can be determined in a robust way, 
in the case of species far from the reference system,
lineage-specific proteins \cite{Koonin:00,Petrov:10,Eddy:20} 
are more likely to be considered
as being unknown. Hereafter, in order to cope with this drawback,
a set of proteomes of species from other (unicellular) eukaryotic lineages
is also analyzed. 

\section*{Methods}

\subsection*{Choice of a reference system}

Unknown proteins are usually defined through the fact that they do not share any significant homology with other known proteins. As a consequence, the status of a given protein may change each time a new proteome 
is unraveled. In the present study, in order to address this issue, unknown proteins are instead defined with respect to a reference system, namely, a set of well-known proteomes as taxonomically diverse as possible.

As a a reference system for eukaryotic proteomes, 36 proteomes were selected as follows,
among the 398 reference proteomes \cite{Uniprot:14} with more than 10,000 proteins available in Uniprot.\footnote{On June 23$^{th}$, 2020.} 
For each of the three better studied eukaryotic kingdoms, namely, metazoa, viridiplantae and fungi, their taxonomic tree, as provided by Uniprot \cite{Uniprot:07}, was scanned down to the node where at least ten taxons with proteomes of more than 10,000 proteins could be found, retaining for each taxon the proteome with the largest number of proteins.  
This protocol yielded 15, 10 and 11 proteomes for metazoa, viridiplantae and fungi, respectively (see Table \ref{Table:references}), corresponding to a total number of 1,174,474 reference sequences. 

\begin{table*}[t!]
 \caption{A reference system for eukaryotic proteomes.
 For each of the three major eukaryotic kingdoms, 
 proteomes were chosen so as to be as
 taxonomically diverse as possible, among those with more
 than 10,000 proteins.}
 \label{Table:references} 
\hskip 0.8 cm
\begin{tabular}{|c|c|c|c|c|}
 \hline
 \multirow{2}{*}{Kingdom} & \multirow{2}{*}{Taxon} & \multirow{2}{*}{Species$^a$} & \multirow{2}{*}{Uniprot Id.} & \multirow{2}{*}{Proteins} \\
                          & &               &    & \\
 \hline
 \multirow{15}{*}{Metazoa} & Arthropoda & Portunus trituberculatus & PORTR & 99,420 \\
 & Craniata & Homo sapiens & HUMAN & 75,004 \\
 & Rotifera & Brachionus plicatilis & BRAPC & 52,387 \\
 & Demospongiae & Amphimedon queenslandica & AMPQE & 43,437 \\
 & Nematoda & Caenorhabditis japonica & CAEJA & 35,024 \\
 & Brachiopoda & Lingula unguis & LINUN & 34,415 \\
 & Eleutherozoa & Stichopus japonicus & STIJA & 30,032 \\
 & Cephalochordata & Branchiostoma floridae & BRAFL & 28,544 \\
 & Mollusca & Crassostrea gigas & CRAGI & 25,997 \\
 & Anthozoa & Nematostella vectensis & NEMVE & 24,435 \\
 & Annelida & Helobdella robusta & HELRO & 23,328 \\
 & Tunicata & Ciona savignyi & CIOSA & 20,004 \\
 & Trematoda & Opisthorchis felineus & OPIFE & 18,330 \\
 & Tardigrada & Hypsibius dujardini & HYPDU & 14,867 \\
 & Cestoda & Hydatigena taeniaeformis & HYDTA & 11,591 \\
 \hline
 \multirow{10}{*}{Viridiplantae} & Poaceae & Aegilops tauschii & AEGTS & 214,162 \\
 & Musaceae & Ensete ventricosum & ENSVE & 58,382 \\
 & Papaveraceae & Papaver somniferum & PAPSO & 41,351 \\
 & Pentapetalae & Arabidopsis thaliana & ARATH & 39,353 \\
 & Coryphoideae & Phoenix dactylifera & PHODC & 34,033 \\
 & Nelumbonaceae & Nelumbo nucifera & NELNU & 31,582 \\
 & Funariaceae & Physcomitrella patens & PHYPA & 30,858 \\
 & Amborellaceae & Amborella trichopoda & AMBTC & 27,371 \\
 & Asparagaceae & Asparagus officinalis & ASPOF & 24,059 \\
 & Bromeliaceae & Ananas comosus & ANACO & 23,408 \\
 \hline
 \multirow{11}{*}{Fungi} & Agaricomycetes & Armillaria gallica & ARMGA & 25,522 \\
 & Blastocladiaceae & Allomyces macrogynus & ALLM3 & 19,092 \\
 & Pezizomycetes & Ascobolus immersus$^b$ & ASCIM & 17,778 \\
 & Dothideomycetes & Corynespora cassiicola$^b$ & CORCC & 17,125 \\
 & Mucorineae & Rhizopus delemar & RHIO9 & 16,971 \\
 & Cunninghamellaceae & Absidia glauca & ABSGL & 14,825 \\
 & Neocallimastigaceae & Piromyces sp. & PIRSE & 14,606 \\
 & Eurotiomycetes & Penicillium camemberti$^b$ & PENCA & 14,390 \\
 & Sordariomycetes & Fusarium poae$^b$ & FUSPO & 14,048 \\
 & Leotiomycetes & Monilinia fructicola$^b$ & MONFR & 13,749 \\
 & Syncephalastraceae & Syncephalastrum racemosum & SYNRA & 11,037 \\
 \hline
\end{tabular}

$^a$With the largest proteome of the taxon.\\
$^b$Ascomycota.\\
\end{table*}

\subsection*{Search of homologues}

Homologues in this reference database were looked for using BLAST \cite{Blast:97} version 2.6.0+,
two proteins being assumed to be homologous when
the E-value of their pairwise alignment is 
lower than $10^{-6}$ \cite{Jones:07,Emili:09,Budak:14}. 
Note that, in order to avoid an overestimation of the number of unknown proteins,
due to the filtering of low-entropy segments, that is, of segments of restricted amino-acid composition,
composition-based statistics \cite{Altschul:01} was not considered (-comp\_based\_stats 0).

\subsection*{Evolutionary distance}

The evolutionary distance between two species has long been estimated by comparing the sequences found in both species for a given protein, such as myoglobin or cytochrome c \cite{Pauling:65} or, for more distantly related species, highly conserved biomolecules like the ribosomal RNA \cite{ZYang:02,YWLim:07,WegenerParfrey:18}.
With the advent of whole genome sequencing projects,
it is nowadays possible to estimate this distance using a large set of common proteins \cite{Dujon:99,Rensing:07}. 
Hereafter, 
the evolutionary distance between a species and the
36 species of the reference system (Table \ref{Table:references})
is estimated by comparing their ubiquitous proteins, that is, proteins that have homologues in all 36 proteomes of the reference set.\footnote{Most of them are likely
to be "housekeeping" proteins.}  
In practice, for each ubiquitous protein of a given species, the closest protein in the reference database is picked and the percentage of differences in their alignment is recorded, the evolutionary distance being the corresponding average over all ubiquitous proteins of the species. For this measure, ubiquitous proteins with more than ten homologues in the considered proteome were not taken into account. 

\section*{Results}

\begin{figure}[t!]
\includegraphics[width=8.0 cm]{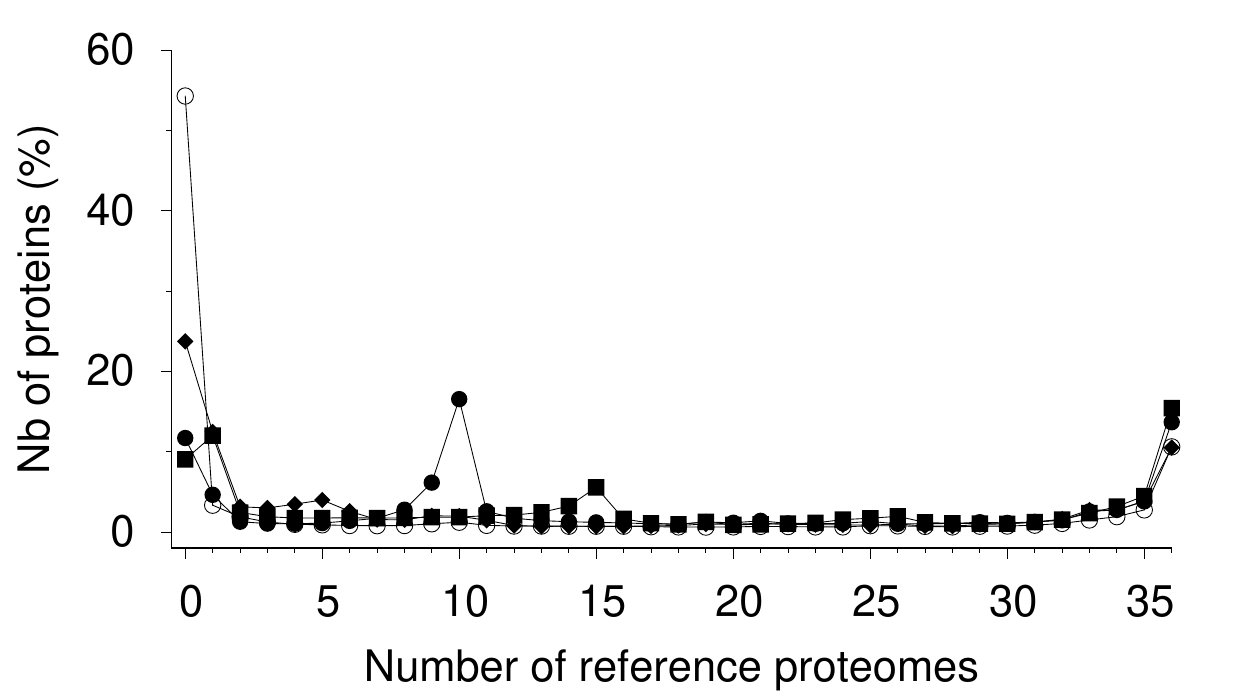}
\caption[]{Percentage of proteins as a function of the number of proteomes of the reference system in which their homologues are found.
Filled squares: proteins from metazoa; filled circles: from viridiplantae; filled diamonds: from fungi; open circles: from other eukaryotes.
}
\label{Fig:proteomes}
\end{figure}

\subsection*{Ubiquitous proteins}

Homologues in the reference database were identified for each protein of the 398 eukaryotic proteomes with more than 10,000 known proteins, that is, 189, 83, 99, 27 proteomes from metazoa, viridiplantae, fungi and other eukaryotic lineages, respectively. On average, whatever the kingdom, 10--15\% of the proteins have homologues in all 36 proteomes 
of the reference set, the largest numbers of them being found in three 
viridiplantae, namely, \textit{Triticum turgidum} (37,911), \textit{Aegilops tauschii} (31,733) and \textit{Hordeum vulgare} (29,499). 
On the other hand, at least 1,000 such ubiquitous proteins were found in all eukaryotic proteomes considered herein, except in the 
cases of \textit{Megaselia scalaris} (864 of them) and
\textit{Eimeria mitis} (336), the later being an apicomplexan parasite \cite{Blake:15}, 
which probably relies on its host (\textit{Gallus gallus}) for compensating the lack of missing ones.  

Interestingly, as suggested by Figure \ref{Fig:proteomes}, a significant number of proteins from metazoa and viridiplantae are kingdom--specific \cite{Gilmartin:98,Rattray:07}, that is, they only have homologues in the proteomes of the reference system coming from their own kingdom (15 and 10 proteomes, respectively). 
On the other hand, fungi do not seem to have a significant number of them (no peak in the case of eleven proteomes), as if their functional diversity were higher. Note however that there is a peak for five proteomes, due to the five ascomycota species of the reference system (see Figure \ref{Fig:proteomes} and Table \ref{Table:references}).

\begin{figure}[t]
\includegraphics[width=8.0 cm]{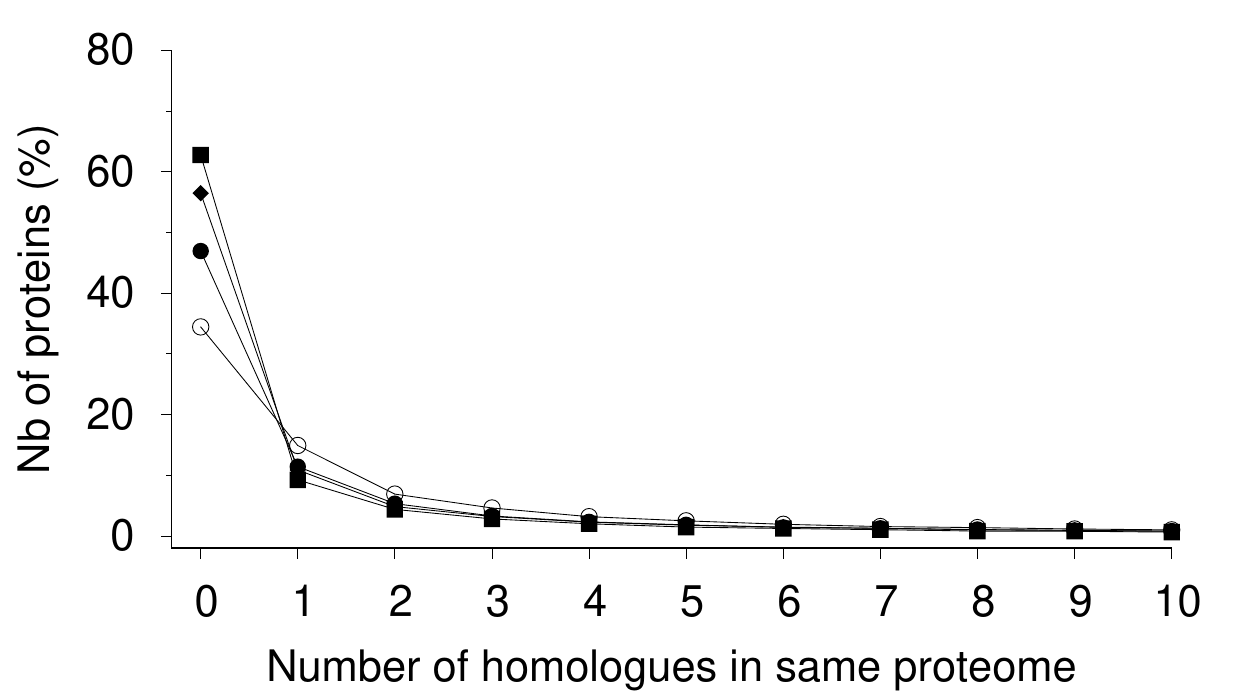}
\caption[]{Number of homologues in their own proteome of unknown proteins, that is, those with no homologue in the 36 proteomes of the reference system.
Filled squares: metazoa; filled circles: viridiplantae; filled diamonds: fungi; open circles: other eukaryotes.
}
\label{Fig:homologues}
\end{figure}

\subsection*{Unknown proteins and singletons}

As shown in Figure \ref{Fig:proteomes}, the percentage of unknown proteins, that is, of proteins not found in the reference database, is around 10\%, on average, in the case of proteomes from metazoa and viridiplantae, around 25\%, in the case of fungi, and as high as 54\%, in the case of the 27 proteomes from other eukaryotic lineages. This later result makes sense if it is assumed that, in this case, homologues of a large amount of unknown proteins were just missed, as a consequence of their high degree of evolutionary divergence. 

Interestingly, as shown in Figure \ref{Fig:homologues}, 
whatever the kingdom,
roughly half of the unknown proteins have homologues within
their own proteome. Note that such proteins are likely to be older than unknown proteins that 
have none, hereafter called singletons. 

\begin{figure*}[t!]
\hskip 0.3 cm
\centering
\includegraphics[width=12.0 cm]{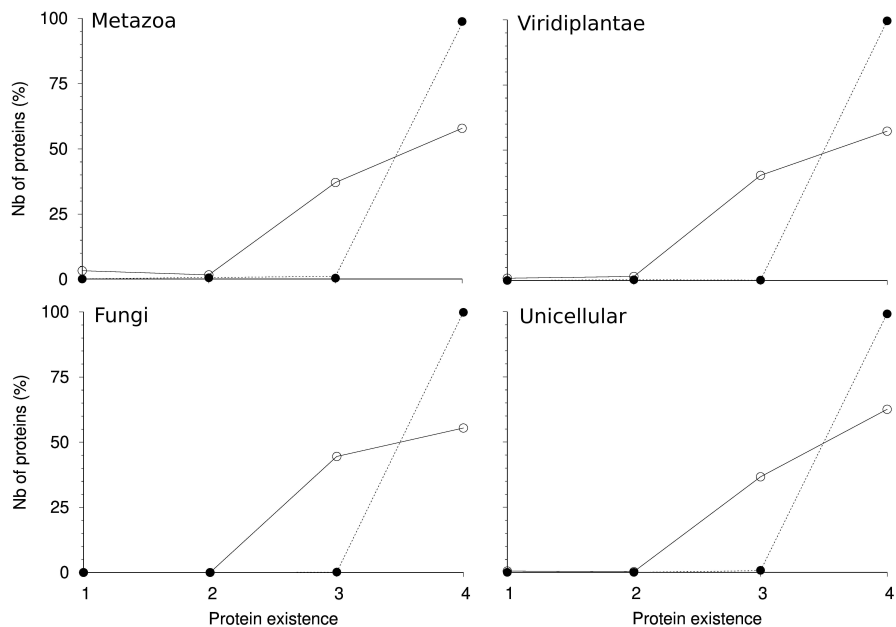}
\caption[]{Number of proteins as a function of their degree of existence, according to Uniprot, for each eukaryotic kingdom. 1 means: known at the protein level; 2: at the transcript level; 3: by homology; 4: predicted.
Open circles and plain lines: ubiquitous proteins; dotted lines: unknown proteins; filled circles: singletons.   
}
\label{Fig:existence}
\end{figure*}

\subsection*{Degree of existence}

In Uniprot, the degree of knowledge about a protein (the so-called degree of existence) is quantified through a number ranging between one (known at the protein level) and four (predicted).\footnote{A few proteins are also classified as being uncertain (fifth degree).} As shown in Figure \ref{Fig:existence}, only a minority of proteins of our dataset are well-characterized ones (degree one), even in the case of the ubiquitous proteins of metazoa (top left). Specifically,
no more than 3\% of them, except in the cases of
\textit{Homo sapiens} (92\%),
\textit{Mus musculus} (84\%), \textit{Rattus norvegicus} (61\%), \textit{Drosophila melanogaster} (59\%), \textit{Danio rerio} (41\%) and \textit{Sus scrofa} (40\%), that is, a few model organisms.  
Likewise, more than 5\% of the ubiquitous proteins are known at the transcript level (degree two)
in the cases of 13 metazoa, 4 viridiplantae and 1 species from an unicellular eukaryotic lineage.

However, whatever the eukaryotic kingdom, around 40--50\%, on average, of the ubiquitous proteins are known by homology (degree three), meaning that they belong to known protein families. 

As expected, such results are in sharp contrast with what is observed for unknown and singleton proteins.
In both cases, for a given species,
no more than 12\% of them are known at the protein level.
Actually, more than 1\% of the singletons are known at the protein level
in the cases of eight species \textit{only}, all of them belonging to the metazoan kingdom. 
For singletons that are, according to Uniprot, actually known by homology,
figures are however a bit higher.
As a matter of fact, they represent more than 5\% of the singletons of a proteome
in the cases of four species, namely, 
\textit{Lipotes vexillifer} (16\%), 
\textit{Leptonychotes weddellii} (10\%), 
\textit{Meleagris gallopavo} (6\%),
\textit{Beauveria bassiana} (6\%) and
\textit{Dictyostelium discoideum} (6\%).
However, for the three first ones,
their number of singletons is unusually low (80 at most),
as well as their number of unknown proteins (138 at most),
strongly suggesting that the annotation of these proteomes
is incomplete, being biased towards proteins with already known homologues. 

\begin{figure*}[t]
\includegraphics[width=16.0 cm]{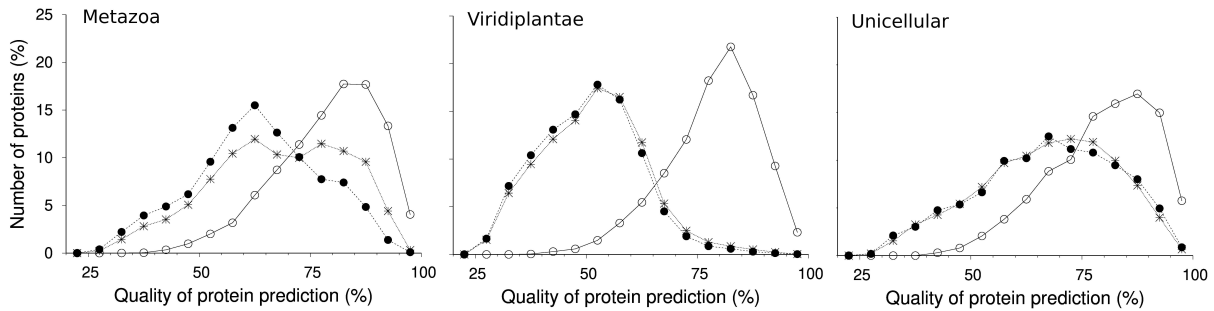}
\caption[]{Number of proteins with a given quality of structural prediction (pLDDT), according to AlphaFold2, for three eukaryotic kingdoms. Over 90\% means: high quality; below 50\%: low one.
Open circles: ubiquitous proteins; stars: unknown proteins; filled circles: singletons.   
}
\label{Fig:af2}
\end{figure*}

\subsection*{Quality of structural prediction}

The predicted tridimensional structures of all proteins of ten proteomes considered in the present study are presently\footnote{As of April 2022.} available in the AlphaFold Protein Structure Database \cite{AF2:22}, namely, two proteomes of the reference system, \textit{Homo sapiens} and \textit{Arabidopsis thaliana}, 
four from other metazoa,
\textit{Mus musculus}, \textit{Rattus norvegicus}, \textit{Drosophila melanogaster} and \textit{Danio rerio},
two from other viridiplantae,
\textit{Oryza sativa} and \textit{Zea mays},
and two from other eukaryotic lineages,
\textit{Dictyostelium discoideum} and \textit{Trypanosoma cruzi}.
Note that there was none from fungi.

In this database, the quality of the prediction of the position of each amino-acid residue of a protein by AlphaFold2\footnote{The second version of AlphaFold.} is provided as a percentage value (coined pLDDT\footnote{Standing for predicted local-distance difference test.}), values over 90\% corresponding to a high quality and values below 50\% to a poor one \cite{AF2:22}.
Herein, the overall quality of the prediction of the structure of a protein is assumed to be given by the average of the quality of the prediction of the position of its residues.

As shown in Figure \ref{Fig:af2}, 
whatever the eukaryotic kingdom,
the overall quality of the prediction of the structures of ubiquitous proteins is rather high, with a median
value of $\approx$ 80\%, values below 50\% being observed in only $\approx$ 1\% of the cases.

Since the predictions of AlphaFold2 are partly based on the information found in the alignment of homologous sequences \cite{AF2:22}, the overall quality of the prediction of the structures of unknown and singleton proteins is expected to be significantly lower. Indeed, whatever the eukaryotic 
kingdom, the median pLDDT value is below 70\%.
In the case of viridiplantae,
it is as low as $\approx$ 50\%,
meaning that the structure of half of the unknown and singleton
proteins of \textit{Oryza sativa} and \textit{Zea mays}
are poorly predicted.  
This may reflect the fact that the proteomes of viridiplantae are less
extensively studied than other eukaryotic proteomes, 
with the consequence that there may have more proteins with not enough known homologues,
that is, with a number of homologues so low\footnote{Below 30.} that it does not allow AlphaFold2 to perform well \cite{AF2:22}.
This may also mean that there are more disordered, hard-to-predict proteins \cite{Pappu:21,Hassabis:21},
among the unknown and singleton
proteins of these two viridiplantae.

At a more general level, note that,
whatever the eukaryotic kingdom,
the quality of the prediction of a structure by AlphaFold2
is similar for unknown and singleton proteins (see Figure \ref{Fig:af2}),
suggesting that the later sequence set is a genuine subset of the former (see also Figure \ref{Fig:existence}).
This is the reason why only singleton proteins are considered hereafter.

Interestingly, AlphaFold2 predicts with great confidence (average pLDDT over 90\%) the tridimensional structure of 192 singletons, among the 13,141 ones (1.5\% of them) found in the AlphaFold Protein Structure Database. 
As suggested by Figure \ref{Fig:proteomes}, only a few (16) come
from viridiplantae, most of them (142) coming from unicellular eukaryotic lineages. In the case of metazoan species, 
accurate structures are predicted for 25 singletons of \textit{Drosophila melanogaster}, 7 of \textit{Danio rerio} and a single one of 
\textit{Mus musculus} and \textit{Rattus norvegicus}.
Among the later 34 cases, according to Uniprot, 7 are known at the protein level, coming all from \textit{Drosophila melanogaster}. 

\subsection*{Singletons with known 3D structure}

If structures predicted with AlphaFold2 were considered above it is, essentially, 
because too few singletons are known at the protein level. As a matter of fact, 
there is a structure in the Protein Data Bank \cite{PDB}
for only 29 of them,
among the 679,509 singletons (0.004\% of them)
found in the 362 eukaryotic proteomes considered. 

Among these 29 singletons with a known tridimensional structure,
seven come from five metazoan species, 
fifteen from two viridiplantae,
and seven from unicellular
eukaryotic lineages.
Amazingly, twelve of them belong to the axoneme
of \textit{Chlamydomonas reinhardtii},
whose structure of the 48-nm repeat is an assembly of 38 different proteins (PDB 6U42),
seven singletons being known as flagellar associated proteins (FAP68, FAP85, FAP95, FAP107, FAP143, FAP222, FAP273).
Two others were identified during the determination
of the structure of the doublet-microtubule, being
not previously associated with cilia (RIB21 and RIB30) \cite{Rzhang:19}.
As expected for genuine singletons, they all seem absent in the axoneme of
an alveolata, \textit{Tetrahymena thermophila} \cite{Winey:22}.
However, for three of them (FAP95, FAP107 and FAP143) 
structural orthologs were found in the cryo-EM electron density map of the axoneme of \textit{Bos taurus} \cite{Brown:21}.
So, these three singletons are likely to have orthologs in the human species, their structure and function being well conserved while their sequences are not.

The ten other singletons from viridiplantae or metazoa with a known tridimensional structure seem also to have a known function, the only obvious orphan being a protein with a CHAD domain found in \textit{Ricinus communis} (PDB 6QV5 \cite{Hothorn:19}).\footnote{Being 76\% identical with an inorganic triphosphatase of a \textit{Duganella} bacterium, this protein is however likely to be a contaminant.} On the other hand, the interleukin 22 of \textit{Danio rerio} (PDB 4O6K \cite{Hartmann:14}) looks like a clean example of a sequence drift quick enough so that homology with its human counterpart is hardly detected at the sequence level, while both display a typical class II cytokine architecture.

\begin{figure}[t]
\includegraphics[width=8.0 cm]{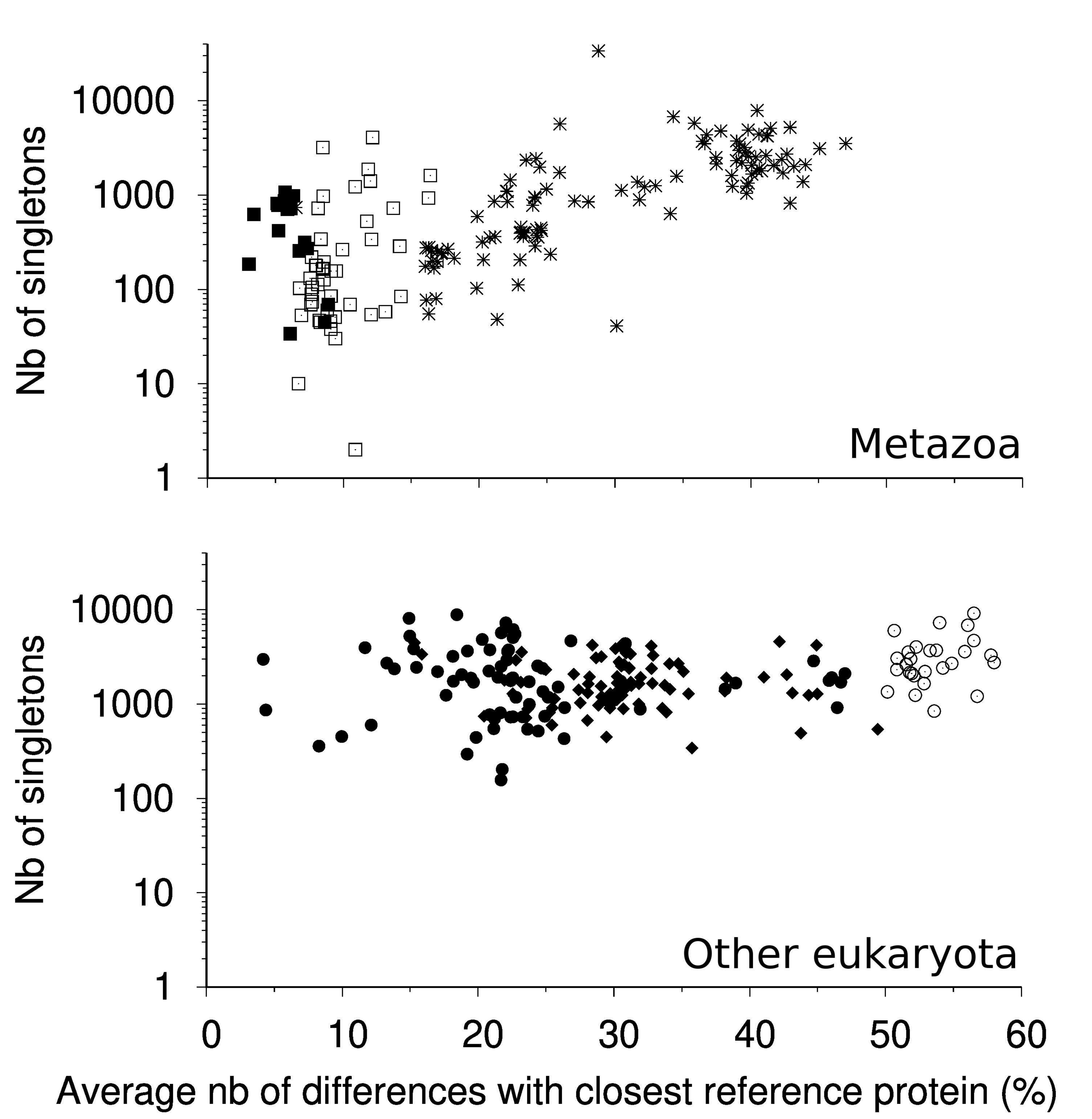}
\caption[]{Number of singletons per proteome, as a function of the evolutionary distance between the species and the reference system. Top, filled squares: primates; open squares: other mammals; stars: other metazoa. 
Bottom, filled circles: viridiplantae; diamonds: fungi; open circles: unicellular eukaryotic lineages.   
}
\label{Fig:evolution}
\end{figure}

\subsection*{Evolution of the number of singletons}

If singletons are randomly added to proteomes from time to time, their number should increase regularly as a function of the evolutionary distance from the reference system. Note that this may not be true for average values since, as mentioned above, proteome annotation may sometimes prove biased towards proteins with known homologues. However, such a trend would be expected for proteomes with the largest number of singletons, at a given evolutionary distance from the reference system.       

As shown in Figure \ref{Fig:evolution} (bottom), no such trend is found
in the cases of viridiplantae and fungi. Indeed, in both kingdoms, the number of singletons per proteome does not seem to vary significantly as a function of the evolutionary distance. Actually, 2,979 and 865 singletons were found in the proteomes of the two species the closest to the reference 
system, namely, \textit{Hordeum vulgare} and \textit{Triticum turgidum}, suggesting that singletons are added to the proteomes of viridiplantae on a short timescale, their proliferation being kept under control afterwards. However, in order to determine on which timescale such a phenomenon occurs, proteomes closer to the reference system need to be considered. 

In the case of metazoa,
an evolutionary trend seems there (Fig. \ref{Fig:evolution}, top),
with at most 1,070 singletons found in the proteomes of
primates (in the case of \textit{Papio anubis}),
and more than 1,000 of them in the proteomes of only eight mammal species, among 65 ones, the largest number of singletons (4,079) being
found in the proteome of \textit{Cricetulus griseus}. 
For other metazoa, the largest number of singletons
is found in the proteome of \textit{Liparis tanakae}, namely, 33,715. 
Note that this is an unusually large number,
almost four times over the largest value found in viridiplantae, namely, 8,875, in the proteome of
\textit{Trifolium medium}.

Of course, the observed trend could be a consequence of the increasing evolutionary distance from the reference system, the homology of more and more proteins being not recognized as a consequence of a too quick neutral sequence drift \cite{Jukes:69,Kimura:83}, like in the case of the interleukin 22 mentioned above. However, since such a drift is not observed for eukaryotic species other than metazoan ones (bottom of Fig. \ref{Fig:evolution}), 
it would anyway mean that singleton dynamics is different in metazoan species.    

Note that while, on average, viridiplantae have more ubiquitous proteins than fungi, namely, 3,453 $\pm$ 3,552 (median value: 2,519) and 849 $\pm$ 351 (median value: 761), respectively, 
their average numbers of singletons are similar, namely, 2,388 $\pm$ 1,958 (median value: 1,774) and 1,922 $\pm$ 1,077 (median value: 1,638), respectively. 
On the other hand, singletons from unicellular
eukaryotic lineages are, on average, more numerous, namely,
3,315 $\pm$ 1,950 (median value: 2,757).
However, 
this is also a likely consequence 
of the high degree of evolutionary divergence
of these proteomes, with respect to the reference system.

\section*{Conclusion}

Defining ubiquitous, unknown and singleton proteins with respect to a set of 36 eukaryotic proteomes as taxonomically diverse as possible, the following results have been obtained.

For instance, a significant number of proteins from metazoa and
viridiplantae are kingdom--specific, that is, they
only have homologues in the proteomes of the reference system coming from their own kingdom.
However, in the case of fungi, such proteins seem rare (Fig. \ref{Fig:proteomes}).  

Noteworthy, roughly half of the unknown proteins are singletons,
that is, no homologue could be found for them, either in the reference system or in their own proteome (Fig. \ref{Fig:homologues}).
On the other hand, there are at least 1,000 ubiquitous proteins in nearly all 398 eukaryotic proteomes considered, the main exception being the proteome of \textit{Eimeria mitis}, an api-complexan parasite.

According to Uniprot, no more than 3\% of the proteins of a given eukaryotic proteome are known at the protein level,
except in the case of the ubiquitous proteins 
of six metazoan species.
As a matter of fact, whatever the eukaryotic kingdom,
most ubiquitous proteins (40--50\% of them) are only
known by homology (Fig. \ref{Fig:existence}).
In the case of singletons, 1\% of them are known
at the protein level in eight
metazoan species \textit{only}. Note that this figure rises up to 12\% (157 among 1342 singletons), in the case of the proteome of \textit{Drosophila melanogaster}, likely as a result of the number of studies dedicated to new genes found in \textit{Drosophila} \cite{Tautz:03,Long:04,Schlotterer:14,Bornberg:21}.

As expected, in the case of ubiquitous proteins,
tridimensional structures are predicted by AlphaFold2 with a high
level of confidence (Fig. \ref{Fig:af2}),
probably because such predictions  
are based on the information found in the alignment
of homologous sequences \cite{AF2:22}.
As a matter of fact,
in the case of singletons, the predictions of AlphaFold2
are poor, in particular in the case of viridiplantae (Fig. \ref{Fig:af2}).

Interestingly, in the case of metazoan species,
the number of singletons seems to increase as
a function of the evolutionary distance 
from the reference system,
as if they were added to their proteomes 
rather regularly (Fig. \ref{Fig:evolution}, top).
On the other hand, no such trend is
found in the cases of viridiplantae or fungi (Fig. \ref{Fig:evolution}, bottom). 
Though this phenomenon needs to be confirmed,
by considering proteomes closer to the reference system, such results suggest that the timescale 
on which singletons are added to proteomes is different
in metazoa and in other eukaryotic kingdoms.
It could also mean that the dominant underlying mechanisms 
are not the same, as already assessed
in the case of fungi \cite{RuizTrillo:22}.


\end{document}